\RequirePackage{fix-cm}
\documentclass[smallextended]{svjour3}       % onecolumn (second format)
\smartqed  % flush right qed marks, e.g. at end of proof
\usepackage{graphicx}
\begin{document}

\title{Geodesic dynamics in Chazy-Curzon spacetimes%\thanks{Grants or other notes
}


\author{F. L. Dubeibe \and J. D. Arias H. \and J. E. Alfonso         
%etc.
}


\institute{F. L. Dubeibe \at
              Grupo de Investigaci\'on Cavendish, Facultad de Ciencias Humanas y de la Educaci\'on, Universidad de los Llanos, Villavicencio 500017, Colombia \\
%              Tel.: +123-45-678910\\
 %            Fax: +123-45-678910\\
              \email{fldubeibem@unal.edu.co}           %  \\
           \and
           J. D. Arias H. \at
              Facultad de Ciencias B\'asicas e Ingenier\'ia, Universidad de los Llanos, Villavicencio 500017, Colombia
              \and
           J. E. Alfonso \at
           Departamento de F\'Isica, Facultad de Ciencias, Universidad de los Andes, Bogot\'a 111711, Colombia
}

\date{Received: date / Accepted: date}

\maketitle

\begin{abstract}
In the last decades, the dynamical studies around compact objects became a subject of active research, partially motivated by the observed differences in the profiles of the gravitational waves depending on the dynamics of the system. In this work, via the Poincar\'e section method, we conduct a thorough numerical analysis of the dynamical behavior of geodesics around Chazy-Curzon metrics. As the main result, we find only regular motions for the geodesics in all cases, which suggest the existence of the so-called Carter's constant in this kind of exact solutions. Moreover, our simulations indicate that in the two-particle Chazy-Curzon solution, some oscillatory motions take place as in the classical MacMillan problem.
\keywords{Nonlinear dynamics and chaos \and general relativity and gravitation \and  exact solutions}
\PACS{04.20.-q \and 05.45.-a \and 04.20.Jb}
\end{abstract}

\section{Introduction}

Unlike in Newton's theory of gravitation, there is no exact solution for a double-star system in orbital motion in general relativity. This is due to the fact that in the strong field regime the spacetime describing binary systems is strongly nonlinear and time-dependent \cite{Poisson2014}. Nowadays, in literature, this problem is tackled in two different ways: by means of approximation methods ({\it e.g.} post-Newtonian approximations) or resorting to numerical relativity \cite{Blanchet2001}. Despite these partial solutions, there is always an imperative need to find exact solutions describing astrophysical scenarios, even if the solutions are not completely realistic. Exact solutions can be used as initial conditions in realistic situations or to fine-tune numerical codes. For example, the Wald's electro-vacuum exact solution in Schwarzschild spacetime \cite{Wald1974} it has been extensively used to test general relativistic force-free codes (see {\it e.g.}  \cite{Paschalidis2013}).

Since the seminal papers of Chazy \cite{Chazy1924} and Curzon \cite{Curzon1925}, there have been many attempts to find a suitable astrophysical exact solution of Einstein and Einstein-Maxwell equations modeling two or more compact objects (see {\it e.g.}. the works of Papapetrou \cite{Papapetrou1945}, Majumdar \cite{Majumdar1947}, Bonnor \cite{Bonnor1966}, Hartle \& Hawking \cite{Hartle1972}, Kramer \& Neugebauer \cite{Kramer1980}, Emparan \cite{Emparan2000}, and Manko {\it et al.} \cite{Manko2014}). In all cases, the equilibrium between the constituents is reached either by electrostatic repulsion or by the so-called Weyl struts. In particular, the two-particle Chazy-Curzon metric comes from the superposition of two Newtonian point sources at different positions on the symmetry axis \cite{Griffiths2009}. Such superposition gives place to a conical singularity which holds the two sources apart in a static configuration \cite{Einstein1936}. At this point, it is important to note that notwithstanding the simplicity and practicality of the Chazy-Curzon metrics, the true nature of the sources has been evaded for decades, \cite{Abdelqader2013} and therefore the dynamics of timelike geodesics around this metrics has never been considered in the literature.

The dynamics of test particles orbiting around two compact objects is of physical interest in at least two contexts. First, it could serve as a first approximation to the problem of the existence (or not) of a third isolating integral of motion in the two-body problem in general relativity. In other words, the numerical evidence of chaotic or regular orbits could shed lights on the existence of the so-called Carter's constant \cite{Carter1968,Dubeibe2007}. Second, previous studies on the dynamics of strong gravity sources have shown that the sensitive dependence in the geodesic dynamics shall be displayed in the gravitational waveforms, that is because the gravitational waveforms can be written in phase space coordinates and hence the differences on nearby trajectories are translated into differences of nearby waveforms \cite{Cornish2001}. Taking into account the previous motivation, in the present paper, we shall study the dynamics around two compact objects using the Chazy-Curzon metric via the Poincar\'e sections method. Given the setup, this study can be considered as a first step and a starting point toward understanding the general relativistic three-body problem.

The paper is organized as follows: In Section \ref{sec:2} we present the solutions derived by Chazy and Curzon along with its most relevant properties. The effective potential and the equations of motion are introduced in  Section \ref{sec:3}. In Section \ref{sec:4}, we briefly analyze the results for the geodesic dynamics in both cases, and finally, in Section \ref{sec:5} we present our conclusions and discuss some possible applications of the results obtained.

%%%%%%%%%%%%%%%%%%%%%%%%
\section{The Chazy-Curzon solution}\label{sec:2}
%%%%%%%%%%%%%%%%%%%%%%%%

For the sake of completeness, we shall begin by deriving the Chazy-Curzon solution \cite{Chazy1924,Curzon1925}. To do so, let us start with the simplest metric for a static axisymmetric spacetime, which was first derived by Weyl \cite{Weyl1917} and reads as 
\begin{equation}\label{eq0}
ds^{2}=-e^{2 \psi} dt^{2}+e^{-2 \psi}[e^{2 \gamma}(d\rho^{2}+dz^{2})+\rho^{2} d\phi^{2}],
\end{equation}
where the metric functions $\psi$ and $\gamma$ are only dependent on Weyl's canonical coordinates $(\rho, z)$. 

With this choice of the metric, Einstein's field equations can be reduced to
\begin{eqnarray}
0&=&\gamma_{,z}-2 \rho \psi_{,\rho} \psi_{,z}, \label{eq1}\\
0&=&\gamma_{,\rho}-\rho (\psi_{,\rho}^{2}-\psi_{,z}^{2}), \label{eq2}\\
0&=&\psi_{,\rho\rho}+\rho^{-1}\psi_{,\rho}+\psi_{,zz}, \label{eq3}
\end{eqnarray}
where equation (\ref{eq3}) is just the classical Laplace's equation in cylindrical coordinates. The simplest non-trivial solution of the overdetermined system of equations (\ref{eq1}-\ref{eq3}) can be calculated by setting the classical expression for the gravitational potential of a point mass source as a solution of (\ref{eq3}), that is
\begin{equation}\label{cc1ppsi}
\psi=-\frac{m}{\sqrt{\rho^{2}+z^{2}}},
\end{equation}
and hence, substituting into Eq. (\ref{eq1}) or (\ref{eq2}), we get 
\begin{equation}\label{cc1pg}
\gamma=-\frac{m^{2} \rho^{2}}{2(\rho^{2}+z^{2})^{2}}.
\end{equation}

Now, it should be noted that due to the linearity of Laplace's equation it is possible to construct a new solution corresponding to the superposition of two Newtonian point sources with masses $m_{1}$ and $m_{2}$ located at $z=a$ y $z=-a$, respectively. In this case, the metric functions are explicitly given as
\begin{equation}\label{cc2pe1}
\psi=-\frac{m_{1}}{\sqrt{\rho^{2}+(z-a)^{2}}}-\frac{m_{2}}{\sqrt{\rho^{2}+(z+a)^{2}}},
\end{equation}
and
\begin{eqnarray}\label{cc2pe2}
\gamma&=&C-\frac{(a^{2}-z^{2}-\rho^{2})m_{1} m_{2}}{2 a^{2}\sqrt{a^{4}+2a^{2}(\rho^{2}-z^{2})
+(\rho^{2}+z^{2})^{2}}} - \frac{\rho^{2}m_{1}^{2}}{2((z-a)^{2}+ \rho^{2})^{2}}\nonumber\\
&-&\frac{\rho^{2}m_{2}^{2}}{2((z+a)^{2}+\rho^{2})^{2}},
\end{eqnarray}
where $C$ is an integration constant that can be determined from the asymptotic flatness condition
\begin{equation}
\lim\limits_{\rho,z\rightarrow\infty}\psi=0,\quad {\rm and} \quad \lim\limits_{\rho,z\rightarrow\infty}\gamma=0,
\end{equation}
and is given by 
\begin{equation}
C=- \frac{m_{1}m_{2}}{2a^{2}}.
\end{equation}
On the other hand, from the elementary flatness condition, $\lim\limits_{\rho\rightarrow0} \gamma=0$, we get
\begin{equation}
\frac{m_{1}m_{2}}{2a^{2}}\left[\frac{(z^{2}-a^{2})}{|z^{2}-a^{2}|}-1\right]=0,
\end{equation}
which is satisfied if and only if $z>a$ and $z<-a$, while a singularity takes place in the interval $-a<z<a$. Such singularity is interpreted as a Weyl strut, which holds the two particles apart and does not exert a gravitational field \cite{Griffiths2009}. The term strut comes from the fact that, at the lowest order, the stress is approximately equal to $m_1 m_2/4 a^2$, as expected from Newtonian theory.\footnote{At this point, it is important to note that such strut can be removed by adding spin to the sources, see {\it e.g.} \cite{Dietz1982} and \cite{Hernandez1993}, however, these metrics will be considered further in a forthcoming paper.}

The Chazy-Curzon solution can be interpreted in terms of its Newtonian limit through the multipole moments. The knowledge of multipole moments let us infer the physical meaning of each parameter in the solution, and hence, get an idea of the kind of source represented by the particular solution. Although there are many methods for finding multipole moments, the Hoenselaers \& Perj\'es method \cite{Hoenselaers1990}, along with the corrections introduced by Sotiriou \& Apostolatos \cite{Sotiriou2004}, allows us to calculate all moments in an efficient and accurate manner. 

Following the Hoenselaers \& Perj\'es procedure, we find that the first five multipolar gravitational moments $P_n$ for a single particle solution (\ref{cc1ppsi},\ref{cc1pg}) are  
\begin{eqnarray}
&& P_0 = m,\,\, P_1 = 0,\,\, P_2 = -m^{3}/3,\,\, P_3=0, \quad {\rm and} \quad P_4=19 m^{5}/105,
\end{eqnarray} 
whence it follows that the parameter $m$, denotes the total mass, the total angular momentum is zero, and all higher-order mass moments are proportional to increasing powers of $m$, {\it i.e.}, we may infer that the solution describes a static non-spherical source. 

On the other hand, the first five multipolar gravitational moments for the two-particle solution (\ref{cc2pe1},\ref{cc2pe2}) with $m_1=m_2=m$, read as 
\begin{eqnarray}
&& P_0 = 2m,\,\, P_1 = 0,\,\, P_2 = 2 a^2 m-8m^{3}/3,\,\, P_3=0, \,\, {\rm and} \nonumber\\ 
&& P_4=2 a^4 m-64 a^2 m^3/7+608 m^{5}/105, 
\end{eqnarray} 
such that the total mass of the source is $2 m$, the total angular momentum is zero with $P_{i}=0$, for $i$ even, and all higher mass moments are proportional to increasing powers of $m$ and $a$. It means that the two-particle solution represents a static pair of non-spherical sources separated by a fixed distance $2 a$.

%%%%%%%%%%%%%%%%%%%%%%%%
\section{Effective potential and equations of motion}\label{sec:3}
%%%%%%%%%%%%%%%%%%%%%%%%

In general, the dynamics of a test particle in General Relativity is determined by the space-time curvature, {\it i.e.}, from the metric. The test particles can exhibit in general two types of motion: bounded and unbounded orbits and, as in the Newtonian orbital dynamics, each type of motion is solely specified by means of the effective potential. Such potential can be derived as follows (see {\it e.g.}  \cite{Dubeibe2016}): {\it (i)} From (\ref{eq0}) and the relation $2{\cal L}=g_{\mu\nu}\dot{x}^{\mu} \dot{x}^{\nu}$, we get 
\begin{equation}\label{Lag}
2{\cal L}=e^{-2\psi}\left[e^{2\gamma}\left(\dot{\rho}^2 + \dot{z}^2\right)+ \rho^2 \dot{\phi}^2\right]- e^{2\psi} \dot{t}^2,
\end{equation}
with ${\cal L}$ the relativistic Lagrangian and $\dot{x}^{\mu}$ the four-velocity of the test particle.
{\it (ii)} From the cyclic coordinates $t$ and $\phi$, the associated four-velocities read as
\begin{equation}\label{EL}
\dot{t}=e^{-2\psi} E,\quad \dot{\phi}= \frac{e^{2\psi} L}{\rho^{2}},
\end{equation}
where the constants $E$ and $L$ are related to the Killing vectors $\xi_{t}$ and $\xi_{\phi}$, representing the total energy and the angular momentum of the test particle, respectively. {\it (iii)} For test particles, the Lagrangian satisfies the relation $2{\cal L}=-\delta$, with $\delta=1$ for massive particles. {\it (iv)} Using the previous relations and some straightforward algebra, the effective potential can be defined as
\begin{equation}\label{potef}
\Phi(\rho,z)=e^{-2\gamma } \left(E^2-\frac{e^{4\psi} L^2}{\rho ^2}-e^{2\psi}\right).
\end{equation}
It should also be noted that, from the definition of $\Phi$, the motion is restricted to regions where $\Phi(\rho,z)\geq 0$. 

On the other hand, the equations of motion resulting from the Lagrangian formalism can be written as
\begin{eqnarray}
&&\ddot{\rho}+(\dot{\rho}^{2}-\dot{z}^{2})(\gamma_{,\rho}-\psi_{,\rho})+2\dot{\rho}\dot{z}(\gamma_{,z}-\psi_{,z})+e^{-2\gamma}\left[E^2 \psi_{,\rho}+(\rho \psi_{,\rho}-1)\frac{L^{2}e^{4\psi}}{\rho^3}\right]=0, \nonumber\\ 
&&\ddot{z}-(\dot{\rho}^{2}-\dot{z}^{2})(\gamma_{,z}-\psi_{,z})+2\dot{\rho}\dot{z}(\gamma_{,\rho}-\psi_{,\rho})+e^{-2\gamma}\psi_{,z}\left[E^2 +\frac{L^{2}e^{4\psi}}{\rho^2}\right]=0,\label{Ecmovfi}
\end{eqnarray}
while, in the case of equatorial motion in axisymmetric static spacetimes, the equation of motion reduces to
\begin{eqnarray}\label{Ecmovecu}
&&\ddot{\rho}+\dot{\rho}^{2}(\gamma_{,\rho}-\psi_{,\rho})+e^{2(2\psi-\gamma)}(\rho \psi_{,\rho}-1)\frac{L^{2}}{\rho^3}+e^{-2\gamma} E^2 \psi_{,\rho}=0.
\end{eqnarray}

%%%%%%%%%%%%%%%%%%%%%%%%
\section{Results and discussions}\label{sec:4}
%%%%%%%%%%%%%%%%%%%%%%%%

%%%%%%%%%%%%%%%%%%%%%%%%
\subsection{Case 1: One-particle solution}
%%%%%%%%%%%%%%%%%%%%%%%%

\begin{figure}%[h]A
\centering
\includegraphics[scale=0.55]{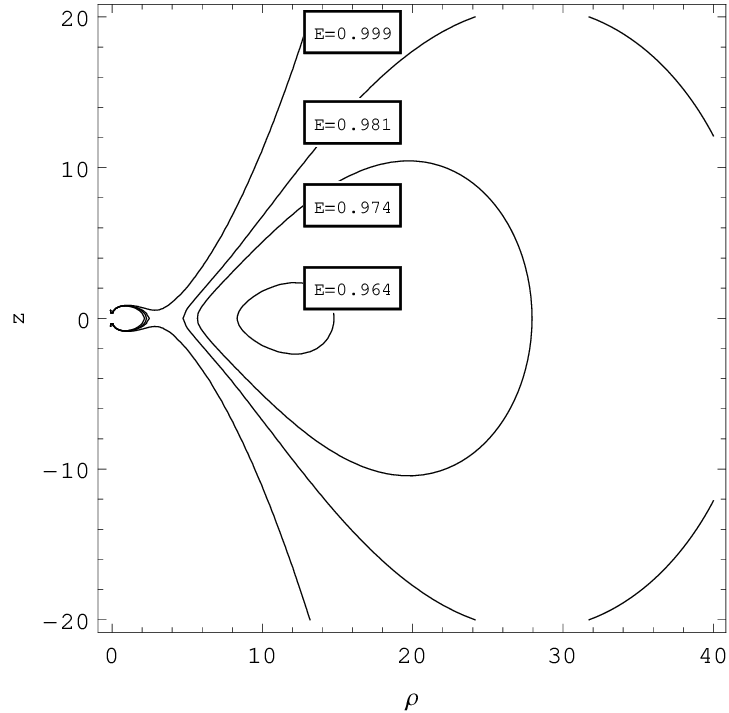}
\caption{Contour plot of the effective potential (\ref{potef}) using the Chazy-Curzon solution (\ref{cc1ppsi},\ref{cc1pg}) with $L = 4.0$ and different values of energy $E$.}
\label{fig1}
\end{figure}

Let us start by considering the motion of a test particle in presence of a single Chazy-Curzon source, such that the metric functions are given by (\ref{cc1ppsi}) and (\ref{cc1pg}) with $m=1$. A typical contour plot of the respective effective potential (\ref{potef}) is shown in Fig. \ref{fig1} for $L=4$ and different values of $E$.  The main difference with the case of fixed $E$ is that as the value of $L$ increases, the confinement region moves to the $+\rho$ direction. As can be noted from Fig. \ref{fig1}, there are three feasible types of trajectories: bounded, unbounded and plunging orbits. 

\begin{figure}%[h]A
\centering
\includegraphics[scale=0.55]{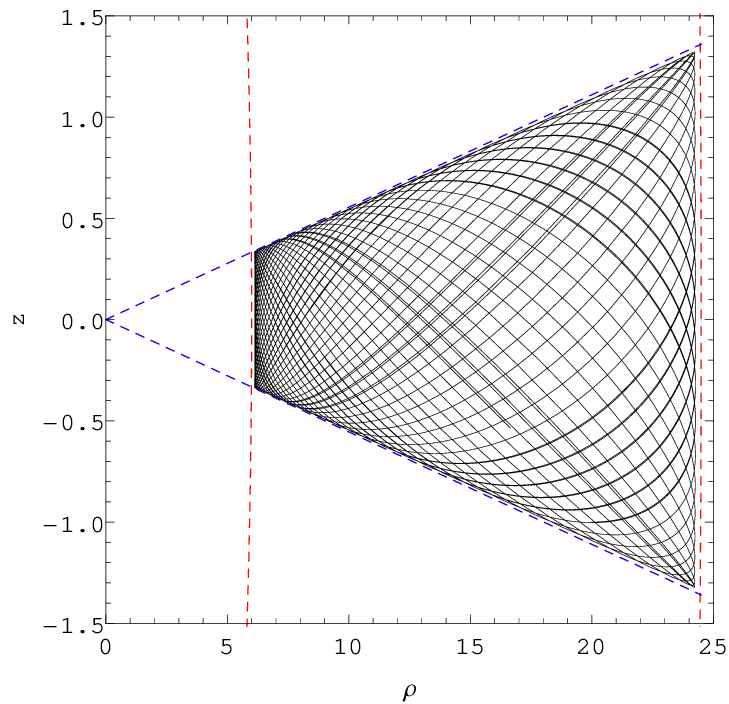}
\caption{Typical bounded orbit in Weyl cylindrical coordinates for $L = 4.0$, $E= 0.972, \rho_{0} = 20, z_{0} = -1, \dot{z}_{0}= 0, \dot{\rho}_{0}= 0.08$, and $\phi_{0} = \pi/6$.}
\label{fig2}
\end{figure}

In Fig. \ref{fig2}, a confined orbit is presented for initial conditions inside a closed contour of the effective potential, with initial conditions and parameters $L = 4, E = 0.472, \rho_{0} = 20, z_{0} = -1, \dot{z}_0 = 0, \dot{\rho}_0= 0.08$ and $\phi_{0} = \pi/6$. The periodicity of this orbit will be determined later through the study of the Poincar\'e sections. It should be noted that when setting the value of the initial condition on the $z$-coordinate $z_0 = 0$ with $\dot{z}_0=0$, the motion is restricted to the equatorial plane. 

An additional zone of possible motion is the one that surrounds the origin (see Fig. \ref{fig1}). For test particles inside this region, we observed that their orbits approach the source in an unusual way: if a test particle moves toward the source along the $z=0$ axis it falls directly to the origin, but if it is not the case, the test particle is repelled and then moves to the origin along the $\rho=0$ axis. A detailed analysis of this behavior was carried out by Gautrean \& Anderson who after computing the Kretschmann scalar $\alpha=R_{\mu\nu\sigma\tau}R^{\mu\nu\sigma\tau}$, showed that the singular behavior of this invariant quantity depends on the approaching direction to the origin \cite{Gautrean1967}. 

For the sake of completeness, we have reproduced the calculation of the Kretschmann scalar, obtaining  
\begin{equation}\label{escalar}
\alpha=\frac{16\ m^{2}\ e^{2m\left[\frac{m\rho^{2}}{(\rho^{2}+z^{2})^{2}}-\frac{2}{\sqrt{\rho^{2}+z^{2}}}\right]}}{(\rho^{2}+z^{2})^{6}} \hat{A}
\end{equation}
with 
\begin{eqnarray}
\hat{A}&=&m^{4}\rho^{2}-3m^{3}\rho^{2}\sqrt{\rho^{2}+z^{2}}-6m(\rho^{2}+z^{2})^{5/2}+3(\rho^{2}+z^{2})^{3}\nonumber\\
&+&3m^{2}(\rho^{2}+z^{2})(2\rho^{2}+z^{2}).
\end{eqnarray}

\begin{figure}%[h]A
\centering
\includegraphics[scale=0.45]{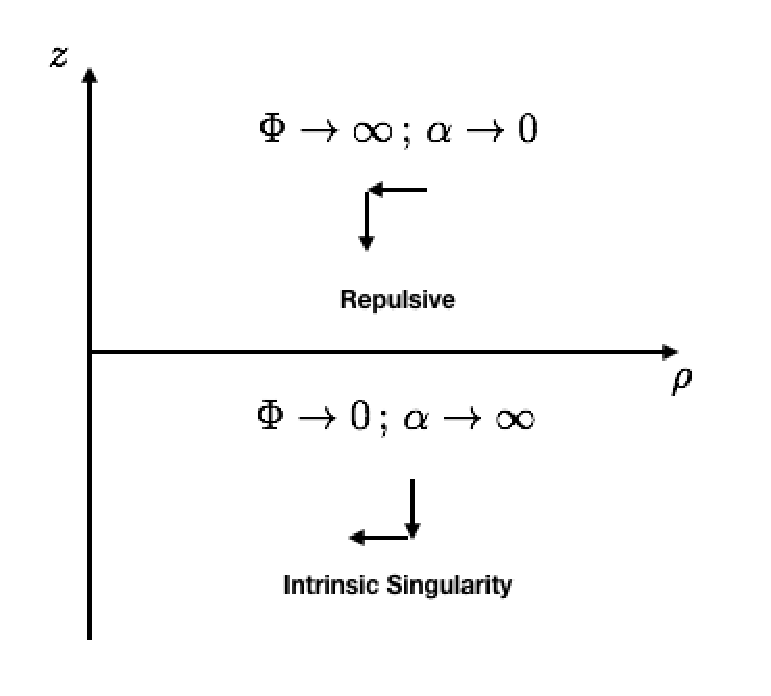}
\caption{Sketch of the behavior of the effective potential $\Phi(\rho,z)$ and the Kretschmann scalar $\alpha$ using two different directions of approach to the origin, {\it (i)} when taking the limit $\rho\rightarrow 0$ and then $z\rightarrow 0$ (upper quadrant) and {\it (ii)} when taking the limit $z\rightarrow 0$ and then $\rho\rightarrow 0$ (lower quadrant).}
\label{fig3}
\end{figure}

A sketch of the behavior of the effective potential $\Phi(\rho,z)$ and the Kretschmann scalar $\alpha$ as we approach the origin along the $\rho$ or $z$-axis, is presented in Fig. \ref{fig3}. In accordance with the results predicted by the geodesic motion, it can be noted that in the first case (upper quadrant of Fig. \ref{fig3}), the Kretschmann scalar $\alpha$ tends to zero, while the effective potential $\Phi(\rho,z)$ tends to infinity, which can be interpreted as a repulsive gravitational phenomenon (see {\it e.g.}  \cite{Herrera2005}); in the second case (lower quadrant of Fig. \ref{fig3}), the Kretschmann scalar $\alpha$ tends to infinity, while the effective potential $\Phi(\rho,z)$ tends to zero, which can be interpreted as an intrinsic singularity. From the previous discussion, it is easy to understand the directional character of the singularity in the one-particle Chazy-Curzon solution.

\begin{figure}%[h]A
\centering
\includegraphics[scale=0.55]{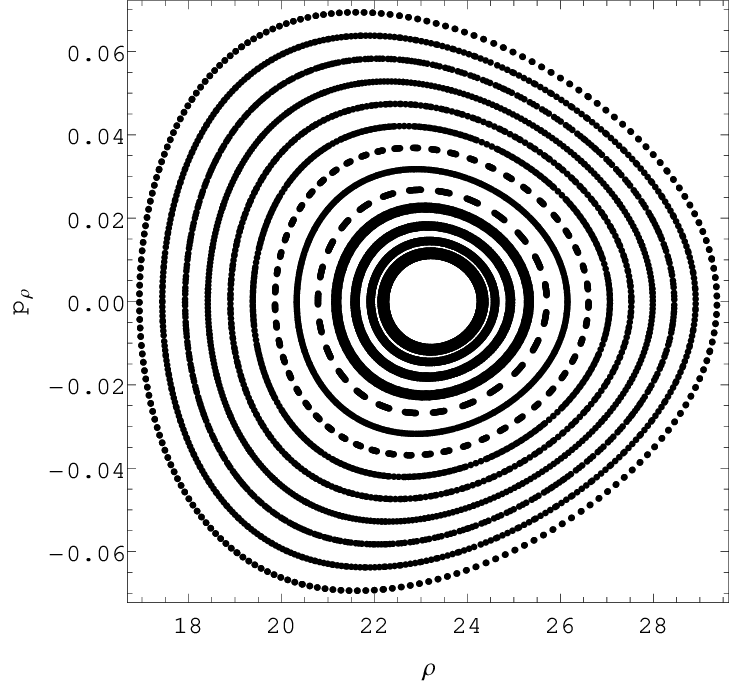}
\caption{Poincar\'e surface of section for the one-particle Chazy-Curzon solution,  for the set of parameters of energy and angular momentum used in Fig. \ref{fig2}.}
\label{fig4}
\end{figure}

In order to fully characterize the geodesic motion of test particles in presence of the one-particle Chazy-Curzon solution, we study the geodesic dynamics by means of the Poincar\'e surfaces of section. In Fig. \ref{fig4} we show the Poincar\'e section $z=0$ in phase space ($\rho, P_{\rho}$) for the set of parameters of energy and angular momentum used in Fig. \ref{fig2}. It can be noted that the available phase space is filled with regular islands and no chaos is observed in the system. The same behavior exhibited in Fig. \ref{fig4}, occurs for 
all the different values of $E$ and $L$ used in this study. Our numerical results suggest the existence of only integrable geodesics.

%%%%%%%%%%%%%%%%%%%%%%%%
\subsection{Case 2: Two-particle solution}
%%%%%%%%%%%%%%%%%%%%%%%%

\begin{figure}%[h]A
\centering
\includegraphics[scale=0.55]{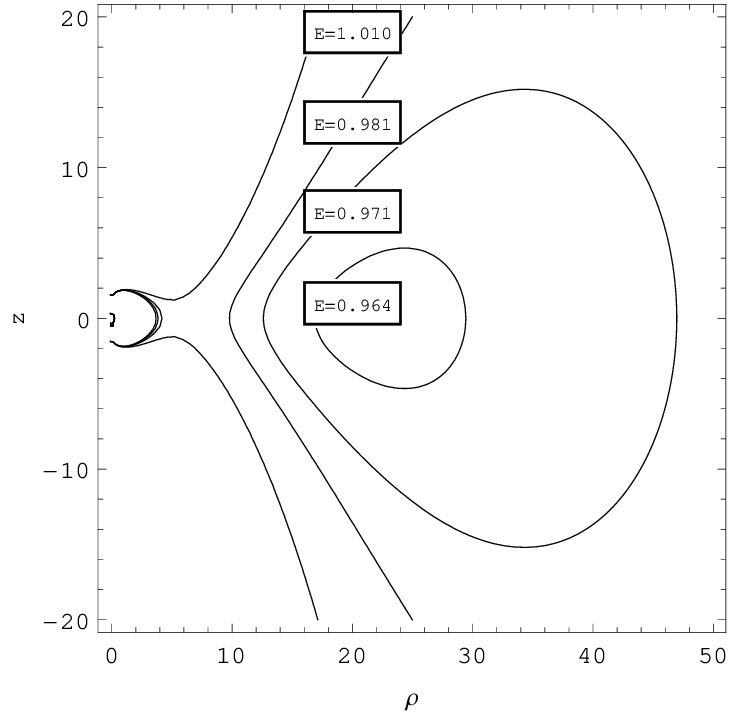}
\caption{Contour plot of the effective potential (\ref{potef}) using the Chazy-Curzon solution (\ref{cc2pe1},\ref{cc2pe2}) with $L = 8.0$ and different values of energy $E$.}
\label{fig5}
\end{figure}

Following the same procedure outlined in the previous subsection, now we consider the motion of test particles in presence of a double Chazy-Curzon source with the metric functions given by (\ref{cc2pe1}) and (\ref{cc2pe2}) and $m_1=m_2=a=1$. In Fig. \ref{fig5} for $L=8$ and using different values of $E$, we show four contour plots of the effective potential (\ref{potef}). Like in the previous case, as the value of $L$ increases, the confinement region moves to the $+\rho$ direction. The structure of the zero velocity surfaces is very similar to the one presented in Fig. \ref{fig1}, except for the fact that in the inner contour there appear two open regions, hence, the three types of trajectories of the one-particle Chazy-Curzon solution are still possible, but, with two available sources to fall in.

\begin{figure}%[h]A
\centering
\includegraphics[scale=0.55]{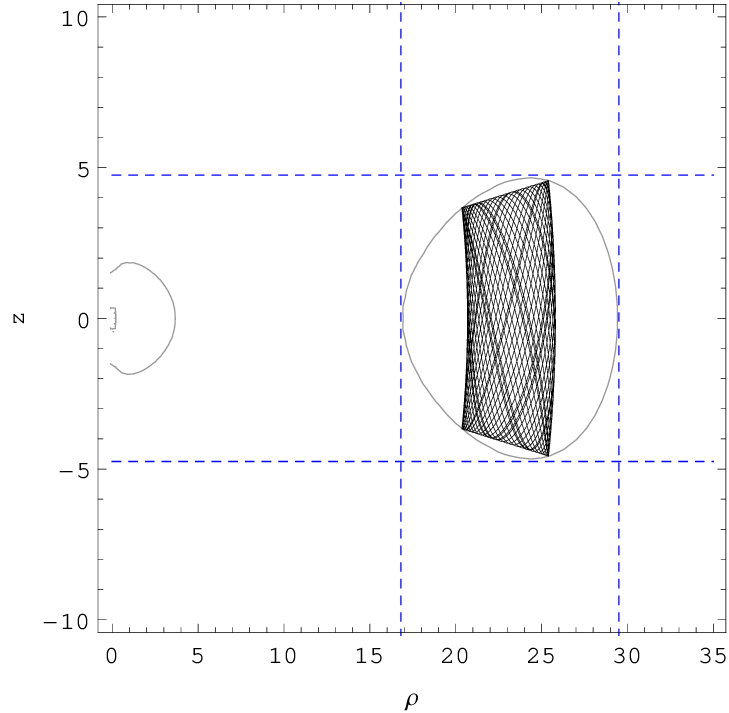}
\caption{Bounded orbit in Weyl cylindrical coordinates for the two-particle Chazy-Curzon solution with $L = 8.0$, $E= 0.964, \rho_{0} = 25, z_{0} = 4.5, \dot{z}_{0}= 0, \dot{\rho}_{0}= 0.012$ and $\phi_{0} = \pi/6$.}
\label{fig6}
\end{figure}

\begin{figure}%[h]A
\centering
\includegraphics[scale=0.55]{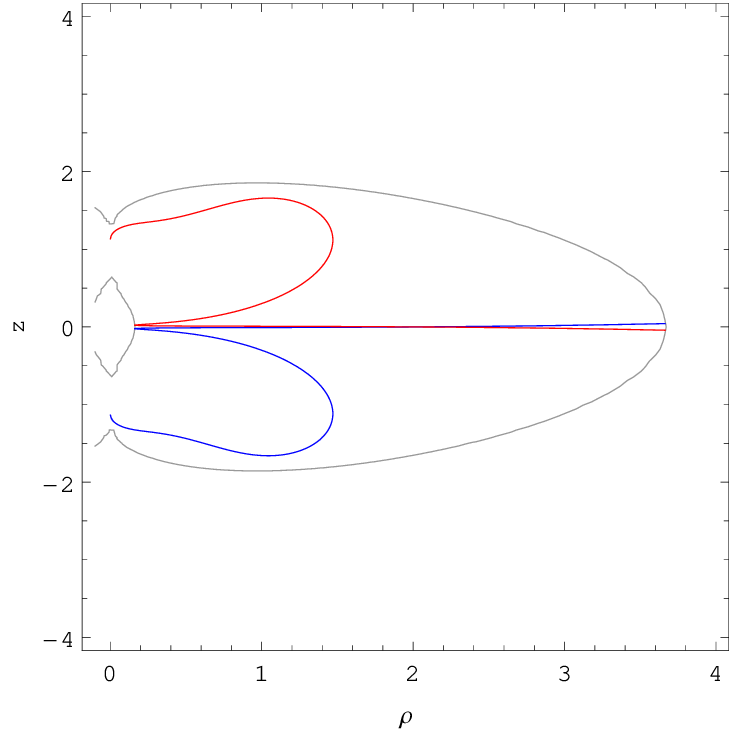}
\caption{Plunging orbits in Weyl cylindrical coordinates for the two-particle Chazy-Curzon solution with $L = 8.0$, $E= 0.964, \rho_{0} = 2, z_{0} = 0, \dot{\rho}_{0}= 0.805$, $\phi_{0} = \pi/6$ and $\dot{z}_{0}= 0.01$ (blue curve) or $\dot{z}_{0}= -0.01$ (red curve).}
\label{fig7}
\end{figure}

\begin{figure}%[h]A
\centering
\includegraphics[scale=0.55]{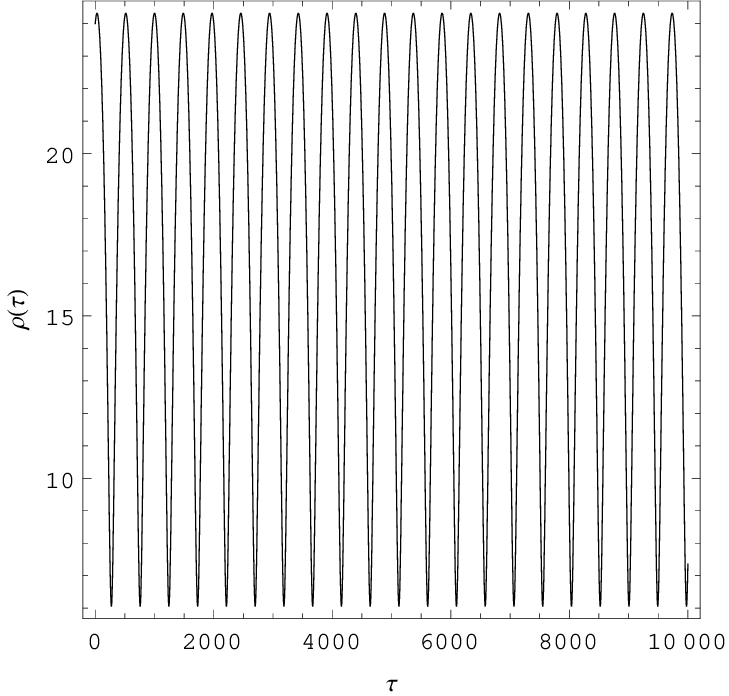}
\caption{Evolution of the $\rho$ coordinate in terms of the affine parameter $\tau$ for test particles. The parameters and initial conditions have been chosen as follows: $E=0.972$, $L=4$, $\rho_0=24, z_0=0$ and $\dot{z}_0=0$.}
\label{fig8}
\end{figure}

As an example of bounded orbit, in Fig. \ref{fig6} we show the trajectory followed by a test particle whose motion is confined to the zero velocity surface presented in the interval $\rho \in (16.8, 29.5)$ and $z \in (-4.75, 4.75)$. In essence, this is the same type of bounded periodic orbit found in the one-particle Chazy-Curzon solution. Additionally, in Fig. \ref{fig7} we show two examples of orbits falling toward the sources. From this figure, it can be seen that there is a reflection symmetry about the $\rho$-axis, which is a consequence of the symmetry properties $(\psi(\rho, z)=\psi(\rho, -z) \land \gamma(\rho, z)=\gamma(\rho, -z))$ of the metric functions. Also, the repulsive character of the Weyl strut can be easily recognized from the same figure. By setting $L = 0$, the repulsive potential between the primaries disappears, and some oscillatory type of motions take place, similar to the ones in the classical MacMillan problem \cite{MacMillan1911} (see Fig. \ref{fig8}).

\begin{figure}%[h]A
\centering
\includegraphics[scale=0.55]{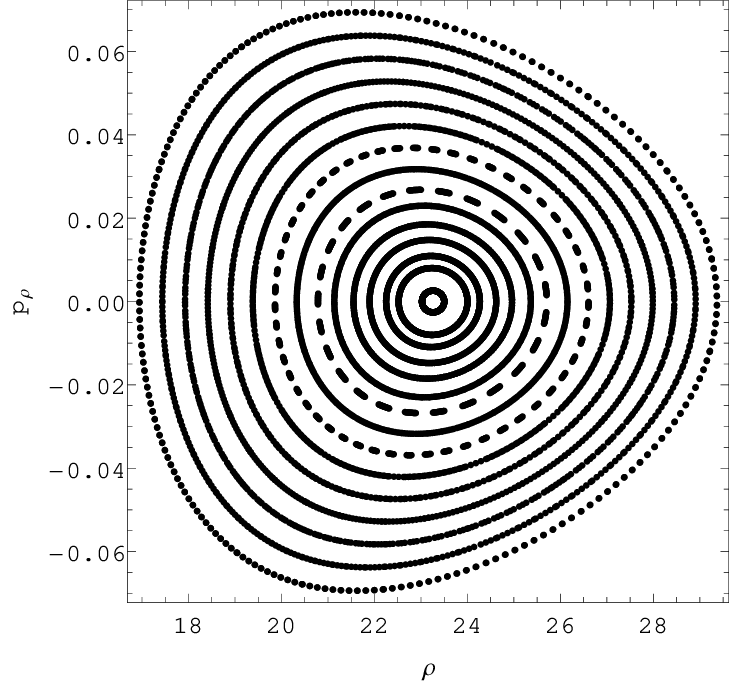}
\caption{Poincar\'e surface of section for the two-particle Chazy-Curzon solution, for the set of parameters of energy and angular momentum used in Fig. \ref{fig6}.}
\label{fig9}
\end{figure}

Concerning the dynamics of the test particles in presence of the two-particle Chazy-Curzon solution, we perform the same analysis using the Poincar\'e surfaces of section. In Fig. \ref{fig9} we show the Poincar\'e section $z=0$ in phase space ($\rho, P_{\rho}$) for the set of parameters of energy and angular momentum used in Fig. \ref{fig6}. From the Kolmogorov-Arnold-Moser (KAM) theory \cite{Holmes1982}, the set of ordered points represent periodic orbits, while closed curves should correspond to the quasi-periodic orbits. The same behavior exhibited in Fig. \ref{fig6}, occurs for the different values of $E$ and $L$ swept in this study, {\it i.e.} almost ten thousand different combinations. As in the case of a single Chazy-Curzon source, our numerical results suggest the existence of only integrable geodesics in the two-particle Chazy-Curzon solution.

%%%%%%%%%%%%%%%%%%%%%
\section{Conclusions}\label{sec:5}
%%%%%%%%%%%%%%%%%%%%%

In the present paper, we have conducted a thorough numerical analysis of the dynamical behavior of geodesics around Chazy-Curzon sources. We start with the one-particle Chazy-Curzon metric and then we move forward to the two particle solution, finding that the multipolar structure suggests that the one-particle Chazy-Curzon solution describes a static non-spherical source, which possesses a directional singularity. On the other hand, the multipolar structure of the two-particle solution shows us that this solution represents a static pair of non-spherical sources separated by a fixed distance $2a$. The two-particle Chazy-Curzon solution satisfies the asymptotic flatness conditions, but the elementary flatness condition is not satisfied in the interval $-a <z<a$, where a conical singularity (Weyl strut) takes place. The strut behaves as a repulsive field in the two-particle Chazy-Curzon solution for $L\neq 0$, but setting $L = 0$, the repulsive potential between the primaries disappears and some oscillatory motions take place, as in the circular Sitnikov problem. 

Once we had an idea of the physical meaning of both solutions, the equations of motion for time-like test particles were derived by means of the Lagrangian formalism. In each case, the system of equations was integrated using a Runge-Kutta-Fehlberg algorithm with an adaptive step size. With this method, the relative error in the system's energy conservation allows us to determine the maximum integration time (in most of the cases $\approx 10^4$) in which the error tolerance is below $10^{-10}$. Given the parameters $E$ and $L$ and the initial conditions  $\rho_{0}, z_{0}, \dot{\rho}_{0}$, the third integral of motion $2{\cal L}=-\delta$ determines $\dot{z}_{0}$. Taking into account that the effective phase-space is three-dimensional, for fixed values of energy and angular momentum, Poincar\'e sections are a convenient tool in dynamical systems theory to analyze the regularity or chaoticity of motion. It is important to note that unlike in the classical, post-Newtonian or Pseudo-Newtonian system (see {\it e.g.}   \cite{Dubeibe2017a,Dubeibe2017b,Zotos2018}), in the general relativistic case the Lyapunov exponents must not be used for the analysis, due to the non-invariance of these exponents in general relativity \cite{Motter2003}.

Concerning the geodesic dynamics, in the present study, we used approximately ten thousand different combinations of values of $E$ and $L$ for each solution. We have scanned the parameter space for energy and angular momentum using the interval $[-1,1]$ for $E$, with a step size of $\Delta E=0.02$, and $[-20,20]$ for $L$, with a step size of $\Delta L=0.4$ (or less if necessary), these intervals define closed contours in the effective potential which is a guaranty for bounded orbits. In both cases (single Chazy-Curzon source and the two-particle Chazy-Curzon solution), the number of initial conditions integrated for each Poincar\'e section is of 50, which is enough to conclude that the phase space is filled by regular orbits. In other words, our findings suggest the existence of the so-called Carter's constant in both systems, which shall provide the fourth conserved quantity necessary to uniquely determine all orbits in each spacetime. We hope our contribution to be useful for a further understanding of the general relativistic three-body problem, in particular, our results could be used as initial conditions in realistic binary simulations or to fine-tune numerical codes.

\begin{acknowledgements}
We thank the anonymous reviewers for their careful reading of our manuscript and their many insightful comments and suggestions. One of the authors (FLD) gratefully acknowledges the financial support provided by Universidad de los Llanos and Colciencias  (Colombia), under Grants No. 8840 and 8863.
\end{acknowledgements}



\end{document}